\documentclass[pra,amsmath,amssymb,twocolumn]{revtex4-2}
 \usepackage{amsmath}
\usepackage{amssymb}
\usepackage{amstext}
\usepackage{amsfonts}
\usepackage{amsxtra}
\usepackage{bm}
\usepackage[usenames]{color}
\usepackage{grffile}
\usepackage{soul}
\usepackage{epstopdf}
\usepackage[dvipsnames]{xcolor}
\usepackage{graphicx}
\usepackage{marvosym}
\usepackage{wasysym}
\usepackage{physics}

\usepackage[colorlinks=true, letterpaper=true, pdfstartview=FitV,
linkcolor=blue, citecolor=blue, urlcolor=blue]{hyperref}

\usepackage[T1]{fontenc}

\begin{document}

\title{Roton-mediated soliton bound states in binary dipolar condensates}

\author{R. M. V. R\"{o}hrs}
\author{R. N. Bisset}
\affiliation{
  Universit\"{a}t Innsbruck, Fakult\"{a}t f\"{u}r Mathematik, 
  Informatik und Physik, Institut f\"{u}r Experimentalphysik, 
  6020 Innsbruck, Austria
}

\begin{abstract}

We investigate the formation of bound states between dark-antidark solitary waves in two-component dipolar Bose-Einstein condensates.
The excitation spectrum contains density and spin branches, and a rotonic feature of the spin branch enables long-range soliton interactions, giving rise to multiple bound states for a single pair, each with a distinct separation.
We show that these bound states originate from periodic modulations of the inter-soliton potential, while individual solitons are surrounded by spatial spin-density oscillations. Both features provide direct signatures of the spin roton.
Collisions between unbound solitons probe this potential, with dipolar interactions enforcing universal bouncing at low velocities, independent of soliton sign, whereas nondipolar solitons may either transmit or bounce.
This distinct behavior offers a realistic path to confirming spin rotons experimentally.

\end{abstract}

\date{\today}
\maketitle

\section{Introduction}

Solitons are localized waveforms that maintain their shape during propagation and after collisions, arising from a balance between nonlinearity and dispersion.
They appear in diverse physical systems, including water waves \cite{Russell1885the,Ablowitz2011nonlinear}, conducting polymers \cite{Heeger1988solitons,Dauxois2006physics}, and nonlinear optics \cite{Kivshar2003optical}, and they play a fundamental role in nonlinear wave phenomena.
Ultracold gases provide an especially versatile platform for studying solitons, offering unparalleled control over experimental conditions \cite{pitaevskii2016bose,kevrekidis2015defocusing}.

Bose-Einstein condensates (BECs) with dipole-dipole interactions promise fundamentally novel possibilities for the study of solitons. 
Condensates made of strongly dipolar atoms, namely chromium \cite{griesmaier2005bose,beaufils2008all}, dysprosium \cite{Lu2011strongly,kadau2016observing} or erbium \cite{aikawa2012bose} are nowadays available in numerous laboratories. 
Even in dilute regimes, the long-ranged and anisotropic interactions give rise to fascinating phenomena---perhaps more familiar in condensed matter settings---including a roton-maxon dispersion relation \cite{odell2003rotons,santos2003roton,blakie2012roton,chomaz2018observation,petter2019probing,schmidt2021roton}, liquid-like droplets \cite{kadau2016observing,Waechtler2016quantum,bisset2016ground,chomaz2016quantum,schmitt2016self} and supersolidity \cite{tanzi2019observation,bottcher2019transient,chomaz2019long}.
When the dispersion relation has a rotonic character, static density oscillations are anticipated to form surrounding a soliton core, and predictions have been made for two dark solitons to form a bound state \cite{pawlowski2015dipolar,Bland2016}.

A parallel research thrust in multicomponent BECs is the study of vector solitons, excitations that span multiple components through intercomponent coupling.
In two-component (binary) mixtures, the excitation spectrum splits into a density branch, corresponding to in-phase density fluctuations, and a spin branch, corresponding to out-of-phase fluctuations.
The spin branch, characterized by fluctuations of the spin density $n_1-n_2$, plays a central role in both the internal structure of vector solitons and their interactions.
An important experimentally realized example in binary condensates is the dark-antidark soliton \cite{hoefer2011dark,yan2012beating,danaila2016vector,farolifi2020observation,chai2020magnetic,Katsimiga2020,mossman2024observation}, a solitary wave in which one miscible component exhibits a phase step and density dip, while the other shows a corresponding density peak.
Of particular interest are bound states of oppositely polarized dark-antidark soliton pairs (also termed dark-dark solitons) \cite{Oehnberg2001dark,charalampidis2016SO2,wang2021dark}, which have been observed experimentally in counter-flowing BECs  \cite{hoefer2011dark,yan2012beating}.
These bound states result from attractive inter-soliton interactions, and give rise to dynamics and stability properties distinct from conventional single-component solitons \cite{rohrs2025magnetic}.
Setting the stage for the present work, we emphasize that although both binary condensates and single-component dipolar condensates can host solitary wave bound states, the underlying physical mechanisms differ. 
Moreover, dipolar mixtures have recently been cooled to quantum degeneracy \cite{Trautmann2018,Ravensbergen2020,chalopin2020probing,Durastante2020,politi2022interspecies,Schafer2023,lecomte2025production,xie2025feshbach,kalia2025creation}, and their capacity to realize vector solitons remains unexplored.

We theoretically investigate dark-antidark solitary waves and their bound states in two-component dipolar BECs confined to quasi-1D geometries.
We compare the density profiles of isolated solitons in nondipolar and dipolar systems.
In the nondipolar case, the soliton width diverges as the spin branch softens, whereas  dipolar interactions generate spin rotons that suppress this broadening and instead induce spatial spin-density oscillations extending far from the core.
We derive the dispersion relations of the uniform system and calculate the interaction potential between two solitons, showing that a spin roton is associated with periodic modulations of the interaction potential, which enable the formation of at least three distinct bound states.
Finally, we study soliton collisions, revealing qualitatively new behavior arising from dipolar interactions.

\section{Formalism}

\subsection{System and interactions}

We study a distinguishable mixture of two BECs in an infinite-tube geometry at the mean-field level.
Each component is labeled by $i\in \{1,2\}$, and atoms in component $i$ at position $\mathbf{x}$ are subject to interactions from atoms in component $j$ at position $\mathbf{x}^\prime$ according to 
\begin{equation}
    \Phi_{ij}^{\rm 3D}(\mathbf{x})= g_{ij}|\Psi_j(\mathbf{x})|^2 + \int d\mathbf{x}\sp{\prime}\,U_{ij}^{\rm 3D}(\mathbf{x}-\mathbf{x}\sp{\prime})\abs{\Psi_j(\mathbf{x}\sp{\prime})}^2,
\end{equation}
where $\Psi_j(\mathbf{x})$ is the wave function of component $j$, and $g_{ij}=4\pi a_{ij}\hbar^2/m$ is the contact coupling constant determined by the $s$-wave scattering lengths $a_{ij}$. The dipolar interactions are described by  
\begin{equation}
U_{ij}^{\rm 3D}(\mathbf{r}) = \frac{3g_{ij}^{\rm dd}}{4\pi}\frac{1-3\cos^2\vartheta}{r^3} \, ,
\end{equation}
with $g_{ij}^{\rm dd}=4\pi a_{ij}^{\rm dd}\hbar^2/m$, where $a_{ij}^{\rm dd} = \mu_0\mu_i\mu_j m/12\pi\hbar^2$ are the dipole lengths associated with magnetic moments $\mu_i$.
For simplicity, we assume both components have the same mass $m$.
The quantity $\vartheta$ represents the angle between the dipole alignment direction $\hat z$ and the vector $\mathbf{r}=\mathbf{x}-\mathbf{x}^\prime$, representing the relative position of the interacting particles.

\subsection{Quasi-one-dimensional GPEs}

The quasi-one-dimensional geometry is realized by a cylindrically symmetric harmonic trap with transverse confinement frequencies $\omega_y=\omega_z\equiv\omega$ and an unconfined axial direction $\omega_x=0$.
In this setting, each condensate wave function separates into a radial and an axial part, the former given by harmonic-oscillator ground states with transverse oscillator length $l=\sqrt{\hbar/m\omega}$, and the latter by $\psi_i(x)$.
After integrating out the transverse degrees of freedom and neglecting the radial zero-point energy, the system reduces to coupled one-dimensional Gross-Pitaevskii equations (GPEs)
\begin{equation}
     i\hbar\partial_t\psi_i=\left(-\frac{\hbar^2}{2m}\partial_x^2+\sum_j\frac{g_{ij} n_j }{2\pi l^2 }   + \sum_j\Phi_{ij} \right)\psi_i\, ,
     \label{Eq:GPE}
\end{equation}
where $n_j(x)=|\psi_j(x)|^2$ is the 1D density and $N_j=\int dx\, n_j(x)$ is the particle number of component $j$. 
The effective 1D dipolar potential $ \Phi_{ij}(x)$ is obtained via a convolution involving the Fourier transformation $\mathcal{F}$ and its inverse $\mathcal{F}^{-1}$ (see \cite{giovanazzi2004instabilities,sinha2007cold,pal2020excitations} for one-component case):
\begin{align}
      \Phi_{ij}(x)&=\mathcal{F}^{-1}\left\{U_{ij}(k)\mathcal{F} \left[\abs{\psi_j}^2 \right]\right\}, \notag \\
U_{ij}(k)&=\frac{g_{ij}^{\rm dd}A_\alpha}{2\pi l^2}\left[1+3Q^2\exp{Q^2}\text{Ei}(-Q^2)\right], \label{Eq:DDIq1D}
\end{align}
with $A_\alpha=[1-3\cos^2\alpha]/2$, $Q=\tfrac{1}{2}k^2l^2$, and $\text{Ei}(x)=-\int_{-x}^\infty dt\, e^{-t}/t$. The angle $\alpha$ specifies the orientation of the dipoles relative to the tube axis $\hat{x}$; here we focus on the case of dipoles aligned perpendicular to the axis, i.e., $\alpha=\pi/2$.

\subsection{Numerics and parameters}\label{Sec:Parameters}
\label{Numerics}
The coupled one-dimensional Gross-Pitaevskii equations (\ref{Eq:GPE}) are solved using a fourth-order Runge-Kutta method on a uniform spatial grid with periodic boundary conditions. 
Following Refs.~\cite{pawlowski2015dipolar,Bland2016,rohrs2025magnetic}, at-rest solitons are prepared by imprinting a $\pi$-phase step in the appropriate component at each time step during imaginary-time evolution.

To isolate the effects of dipole-dipole interactions, we use the same trapping frequencies and average densities in all simulations
(although in Appendix \ref{Appendix:OtherMixtures} we consider other densities).
Specifically, we consider balanced populations $N_1=N_2\equiv N/2$, with average per-component density $n_0=N/2L=667$ atoms $\mu$m$^{-1}$ and a single transverse trapping frequency $\omega=2\pi\times 500$\,Hz.
In the dipolar case, a roton in the spin branch of the dispersion relation can occur over a broad range of parameters but emerges more readily when the components have imbalanced dipole moments \cite{wilson2012roton,lee2021miscibility,kirkby2023spin}. 
As a representative example, we study a binary dipolar condensate based on $^{162}$Dy, with $\mu_1=9.93\mu_{\rm B}$, $\mu_2=0$, and $m=162u$, where $u$ is the atomic mass unit. 
This situation could be experimentally realized using spin mixtures \footnote{Progress has recently been made in creating stable spin mixtures of Dy atoms; see Refs.~\cite{chalopin2020probing,lecomte2025production}.}.
However, our results should be qualitatively applicable across a wide range of dipolar- and nondipolar-dipolar mixtures \cite{Trautmann2018,Ravensbergen2020,chalopin2020probing,Durastante2020,politi2022interspecies,Schafer2023,lecomte2025production,xie2025feshbach,kalia2025creation}, provided a spin roton is present. To demonstrate this generality, we investigate two additional parameter regimes in appendix \ref{Appendix:OtherMixtures}.
For the nondipolar binary condensate, we set $\mu_1 = \mu_2=0$ and take $^{87}$Rb ($m=87u$) as our reference system.

\section{Results}

\subsection{Isolated dark-antidark solitons}

\begin{figure}
	\centering
	\includegraphics[width=0.48\textwidth]{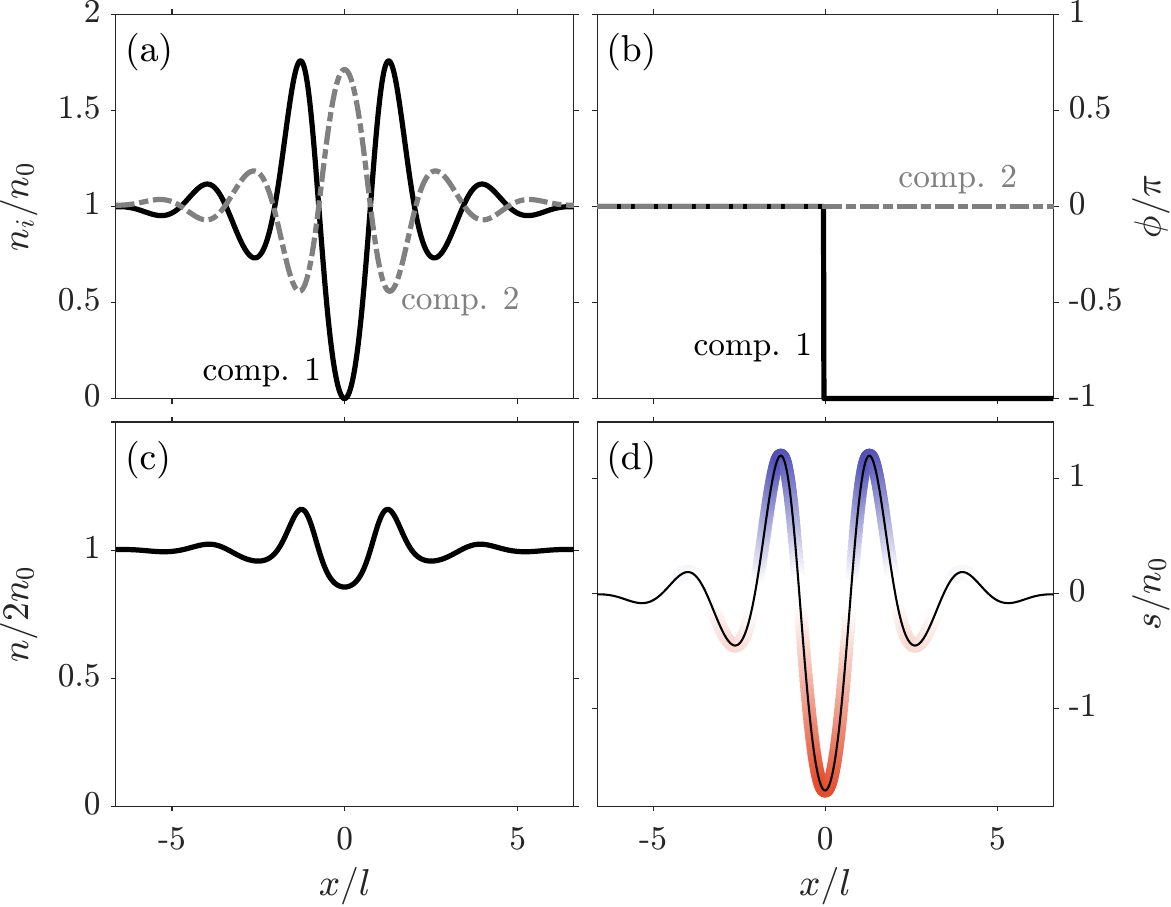}
	\caption{An isolated dark-antidark soliton in a binary dipolar condensate.
	(a) Density profiles: component 1 (solid) and component 2 (dot-dashed).
	(b) Wave function phase profiles.
	(c) Total-density profile.
	(d) Spin-density profile. Colors match the spin densities shown in Figs.~\ref{Fig:Pot}(i-iii) and \ref{Fig:Coll}(b). Scattering lengths: $\{a_{11},a_{12},a_{22}\}/a_0 = \{140,105,140\}$.}
	\label{Fig1:Dens}
\end{figure}

In single-component dipolar BECs, spatial density oscillations arise near boundaries and defects such as hard walls \cite{lu2010spatial}, quantum vortices \cite{yi2006vortex,wilson2008manifestations}, and soliton cores \cite{pawlowski2015dipolar,Bland2016}.
These features are associated with roton modes in the dispersion relation \cite{ronen2007radial,wilson2008manifestations}, analogous to the connection between rotons and spatial density oscillations in superfluid helium \cite{regge1972free,dalfovo1992structure}.
Similarly, Fig.~\ref{Fig1:Dens} shows that in a binary dipolar condensate, a dark-antidark soliton is a localized spin-density perturbation that generates spatial spin-density oscillations around its core, analogous to the density oscillations predicted around a half-quantum vortex \cite{shirley2014half}.

Figure \ref{Fig1:Dens} displays the (a) density and (b) phase distributions of a dark-antidark soliton at rest. The dipolar component (component 1) hosts a dark soliton with a $\pi$ phase step, while the nondipolar component (component 2) forms an antidark structure with uniform phase.
Figure \ref{Fig1:Dens}(c) shows that, while this excitation generates small modulations in the total density $n(x)=n_1(x)+n_2(x)$, Fig.~\ref{Fig1:Dens}(d) reveals pronounced oscillations of the spin density $s(x)=n_1(x)-n_2(x)$.
Since the sign of the spin density also oscillates, we define the soliton sign as the sign of $s(x)$ at its core, determined by which components carries the phase jump.
For the present example, at the core $s(x) < 0$,
which we define as a \textit{negative-core} dark-antidark soliton.
Conversely, when the soliton core exhibits $s(x) > 0$, we refer to it as a \textit{positive-core} dark-antidark soliton.
The characteristic spacing between the resulting peaks and troughs of $s(x)$ can be understood in terms of the system's dispersion relations.

\subsection{Derivation of dispersion relations}

Binary condensates exhibit two gapless Goldstone dispersion relations, arising from the two spontaneously broken gauge symmetries associated with each order parameter.
Because of the intercomponent coupling, one branch is predominantly associated with fluctuations of the total density $n(x)$, while the other corresponds to fluctuations of the
spin density $s(x)$.
Using Bogoliubov theory, we derive an expression for the density ($+$) and spin ($-$) dispersion relation branches (see Appendix \ref{Appendix:BdG}):
\begin{equation}
    \epsilon_{\pm}^2(k)=\frac{\epsilon_1^2+\epsilon_2^2}{2}\pm\frac{1}{2}\sqrt{(\epsilon_1^2-\epsilon_2^2)^2+4G_{12}G_{12}\frac{\hbar^4k^4}{m^2}},
    \label{Eq:BdG}
\end{equation}
where
\begin{equation}
    \epsilon_i^2(k)=\frac{\hbar^2k^2}{2m}\left(\frac{\hbar^2k^2}{2m}+2G_{ii}(k)\right) ,
\end{equation}
with
\begin{align}
    G_{ij}(k)=\sqrt{n_in_j} \left( \frac{g_{ij}}{2\pi l^2}+U_{ij}(k) \right) . \label{Eq:Gij}
\end{align}

\subsection{Role of dispersion relations in soliton structure}

\begin{figure}
	\centering
	\includegraphics[width=0.48\textwidth]{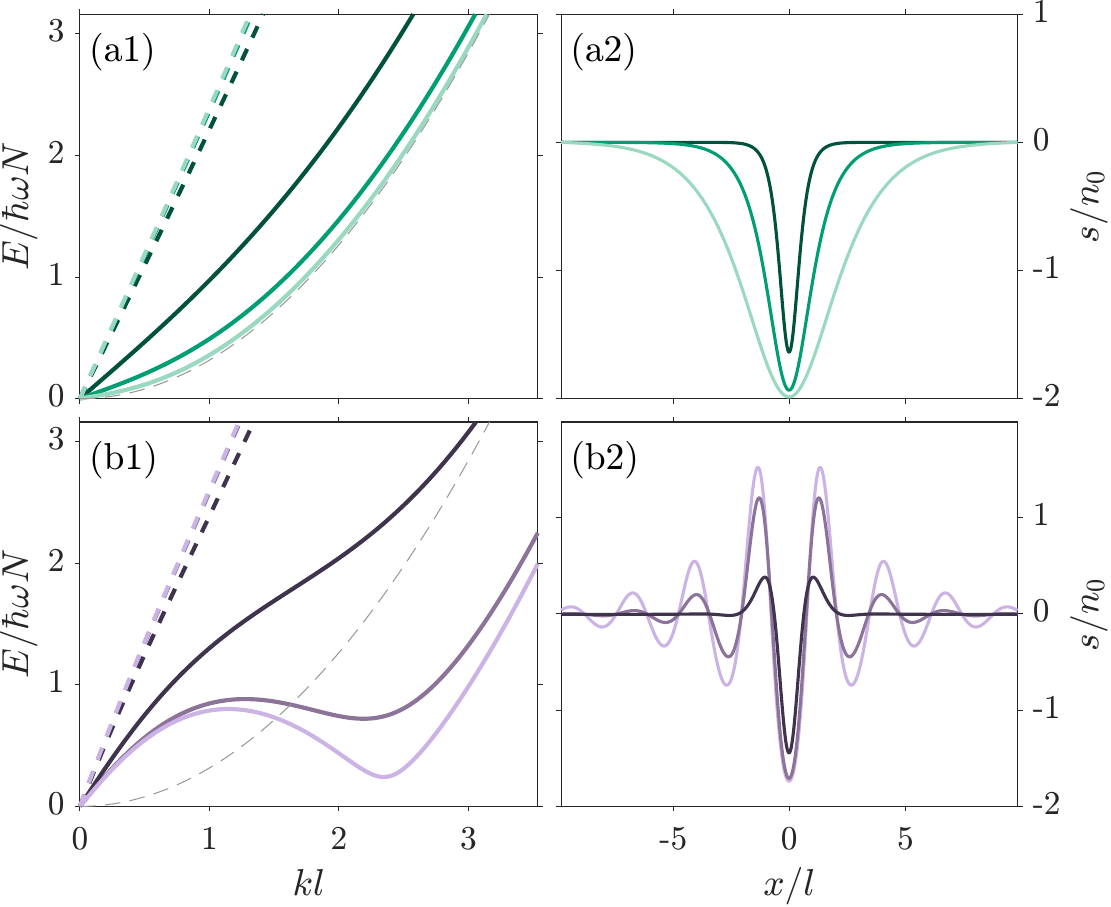}
	\caption{Relationship between the dispersion relations (left) and spin density (right) for a soliton in a binary condensate, shown without (a) and with (b) dipolar interactions.
	(a1) Dispersion relations for nondipolar mixtures with $a_{12} = [70, 95, 99]\,a_0$ from dark to light, and fixed intraspecies interactions $a_{11}=a_{22}=100\,a_0$.
	Solid (dashed) lines denote spin (density) branches; thin dashed line shows free-particle reference.
(a2) Corresponding spin density profiles of dark-antidark solitons. 
(b1) Dispersion relations for binary dipolar condensate with $a_{12} = [70, 105, 108.4]\,a_0$, from dark to light, with $a_{11}=a_{22}=140\,a_0$. 
(b2) Corresponding spin-density profiles of dark-antidark solitons.
}
	\label{Fig2:Dispersion}
\end{figure}

Here we compare the role of the dispersion relations in determining the structure of dark-antidark solitons in dipolar mixtures and contrast this with that of nondipolar mixtures.
Figure \ref{Fig2:Dispersion} shows the dispersion relations [Eq.~(\ref{Eq:BdG})] and the corresponding spin density of isolated solitons for (a) nondipolar and (b) dipolar mixtures for several values of $a_{12}$.
In both cases, the spin branches $\epsilon_-$ (solid lines, shaded from dark to light with increasing $a_{12}$) soften as the miscible-immiscible threshold is approached ($a_{12}^\text{c}=\sqrt{a_{11}a_{22}}=100a_0$ for the nondipolar case and $a_{12}^\text{c}\approx 110a_0$ for the dipolar case).

For nondipolar mixtures [Fig.~\ref{Fig2:Dispersion}(a1)], the spin branch smoothly approaches the free-particle dispersion (thin dashed line), with the spin healing length $\xi_s = \hbar/\sqrt{2mn_0\delta g/\pi l^2}$ diverging at the transition where $\delta g\equiv(\sqrt{g_{11}g_{22}}-g_{12})=0$.
In contrast, the dipolar mixture [Fig.~\ref{Fig2:Dispersion}(b1)] develops a spin-roton minimum \cite{wilson2012roton,lee2021miscibility,kirkby2023spin,kirkby2024excitations}, which deepens with increasing $a_{12}$ and bends the branch below the free-particle (non-interacting) curve, corresponding to effectively repulsive interactions at small $k$ but attractive interactions at large $k$.
This can be understood by noting from Eqs.~(\ref{Eq:BdG})-(\ref{Eq:Gij}) that  deviations from the free particle energy, $\epsilon_0=\hbar^2k^2/2m$, arise from $G_{ij}(k)$, and the changing sign of $\epsilon_-(k)-\epsilon_0(k)$ is caused by the momentum dependence of the dipole-dipole interactions $U_{ij}(k)$
\footnote{
For discussions of the momentum-dependent effective interaction in single-component dipolar gases, see, e.g., Refs.~\cite{santos2003roton,blakie2013depletion,bisset2019static}.
}.
In both systems, the density branches $\epsilon_+$ (dashed lines) depend only weakly on $a_{12}$.

The corresponding spin-density soliton profiles reflect these contrasting dispersion relations \footnote{Note that while the dispersion relation is calculated for the homogeneous system, the numerical simulations introduce a small finite-size effect due to the presence of the soliton. As a result, the critical value of $a_{12}$ is approximately 1\% higher in Figs.~\ref{Fig2:Dispersion}(a2,b2) that in Figs.~\ref{Fig2:Dispersion}(a1,b1)}.
In the nondipolar case [Fig.~\ref{Fig2:Dispersion}(a2)], the soliton width grows with increasing $\xi_s$, while the central spin density approaches $s=-2n_0$, consistent with a magnetic soliton of nearly constant total density \cite{qu2016magnetic}.
In dipolar mixtures [Fig.~\ref{Fig2:Dispersion}(b2)], the roton generates spin-density oscillations of wavelength $\lambda_\text{rot}/l \approx2.7$ around the soliton core. The width of the central minimum remains $\sim\lambda_\text{rot}$, but the decay length of the modulations increases with increasing roton depth.
Note that at the miscible-immiscible threshold, spin-density fluctuations cost zero energy, signally a transition to the alternating-domain supersolid phase \cite{bland2022alternating,li2022long,kirkby2024excitations}.
The presence of spin rotons not only shapes individual solitons but also plays a crucial role in mediating the interactions between two solitons, setting the stage for the formation of multiple bound states.

\subsection{Multiple bound states and soliton-pair interactions }

\begin{figure}[h!]
	\centering
	\includegraphics[width=0.5\textwidth]{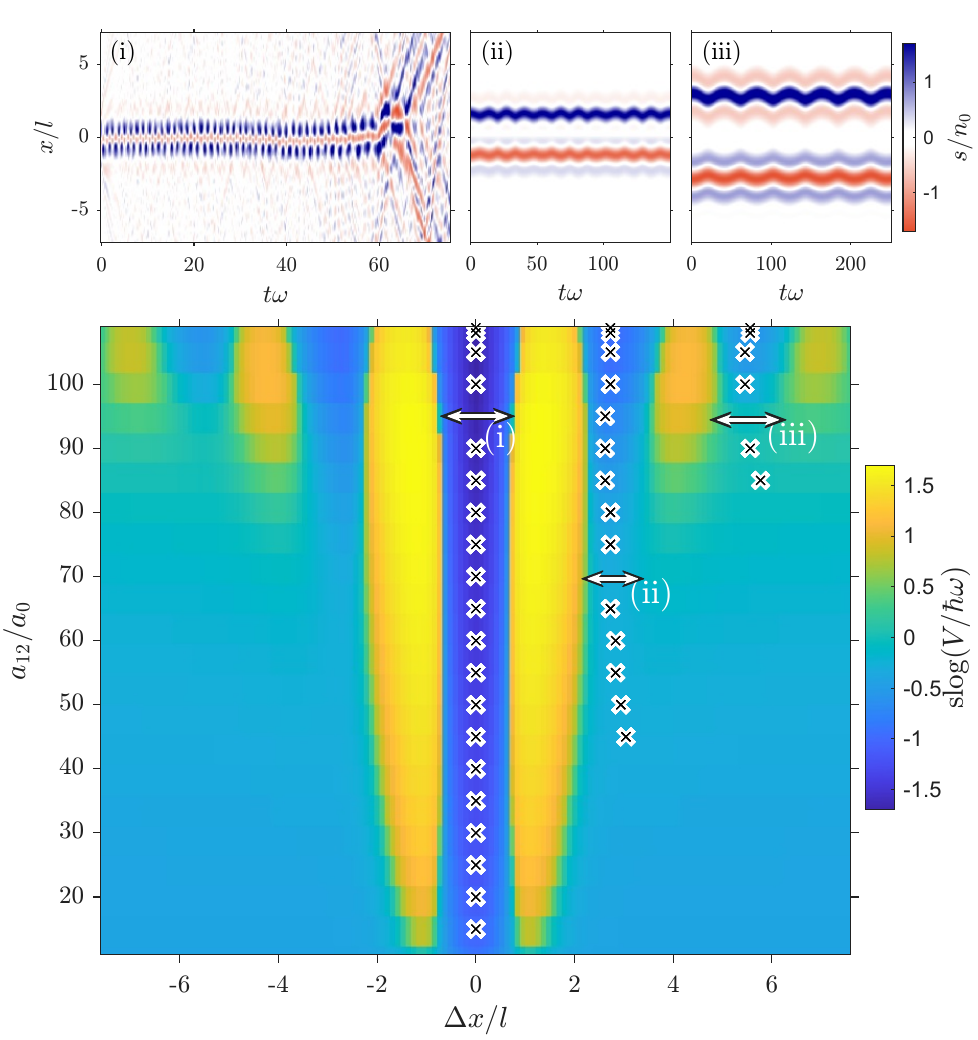}
	\caption{(Main panel) Inter-soliton potential versus separation and interspecies scattering length for binary dipolar condensate.
	Crosses mark local minima where stationary-state bound states exist.
	Panels (i-iii) show oscillations of two solitons about the first, second, and third minima, respectively, for selected $a_{12}$ values (see double-headed arrows), with spin-density color coding as in Fig.~\ref{Fig1:Dens}(d).
	The spin density is shown as a function of position and time.
	Intraspecies scattering lengths are fixed at $a_{11}=a_{22}=140a_0$.
	}
	\label{Fig:Pot}
\end{figure}

The upper panels of Fig.~\ref{Fig:Pot} illustrate representative examples of three types of excited bound states with distinct equilibrium separations.
Panel (i) shows solitons oscillating through one another at $a_{12}/a_0=95$, analogous to magnetic soliton bound states in binary nondipolar condensates \cite{rohrs2025magnetic}.
Panels (ii) and (iii) depict long-range bound states oscillating about equilibrium positions $\Delta x/l \approx 2.7$ and $\Delta x/l \approx 5.4$ for $a_{12}/a_0=70$ and $a_{12}/a_0=95$, respectively, where $\Delta x = x_2-x_1$ with $x_i$ the position of soliton $i$.
Because dipolar interactions break integrability, these structures are not true solitons in the strict sense and radiate energy through phonon emission \cite{pawlowski2015dipolar}.
Such phonons can be seen as fluctuations in the spin density far from the solitons in Fig.~\ref{Fig:Pot}(i), even for early times.
Dissipation-driven acceleration eventually pushes oscillating solitons beyond their escape velocity \cite{pawlowski2015dipolar,bland2017interaction}, as seen in Fig.~\ref{Fig:Pot}(i) at $t\omega \approx 60$.
This counterintuitive behavior indicates that dark-antidark solitons in binary dipolar BECs have negative effective masses, similar to dark solitons in single-component dipolar BECs \cite{pawlowski2015dipolar,Bland2016} and dark-antidark soltions in binary BECs \cite{qu2016magnetic}. We confirm this numerically for our system in Appendix \ref{Appendix:EMS}.

The three different types of bound states with distinct separations can be understood by considering the effective inter-soliton potential
\begin{equation}
V(\Delta x,a_{12}) = \text{sign}(m_\text{sol})\,E(\Delta x,a_{12}),
\label{Eq:ISP}
\end{equation}
which is calculated from the GPEs with
\begin{equation}
E(\Delta x,a_{12}) = E_\text{tot}(\Delta x,a_{12})-E_\text{sep}(a_{12}),
\end{equation}
where $E_\text{tot}$ is the system energy of two zero-velocity solitons separated by a distance $\Delta x$, and $E_\text{sep}$ is the system energy of two well-separated solitons, ensuring $E=0$ at maximal separation.
For dark-antidark solitons, the effective mass satisfies $m_{\rm{sol}}<0$, and its sign is included in Eq.~(\ref{Eq:ISP}) to render the direction of the restoring force for bound-state oscillations in Fig.~\ref{Fig:Pot} more transparent. 

The main panel in Fig.~\ref{Fig:Pot} presents $V(\Delta x, a_{12})$ for two opposite-sign solitons across a broad range of intercomponent scattering lengths $a_{12}$, plotted using a symmetric logarithmic transform,
\begin{equation}
    \text{slog}(x) = \text{sgn}(x)\log_{10}\left(\sqrt{x^2 + 1}\right),
\end{equation}
to highlight small-magnitude features.

The modulation prominence of $V(\Delta x)$---related to the depth of the spin roton---increases up to the roton-instability threshold $a_{12}^\text{c} \approx 110\,a_0$
\footnote{In Fig.~\ref{Fig:Pot}, the highest value of $a_{12}$ considered is at least $1a_0$ below this threshold.}. 
A minimum at $\Delta x = 0$ exists for all $a_{12}$.
As $a_{12}$ increases, additional local minima emerge: one at $|\Delta x|/l \approx 2.7$ for $a_{12} \gtrsim 45\,a_0$, and another at $|\Delta x|/l \approx 5.4$ for $a_{12} \gtrsim 85\,a_0$, resulting in three distinct potential valleys (crosses).
These valleys indicate the distinct separations at which soliton bound states form.

\subsection{Origin of multiple bound states}
\label{Sec:BSOrigin}

\begin{figure}[h!]
	\centering
	\includegraphics[width=0.48\textwidth]{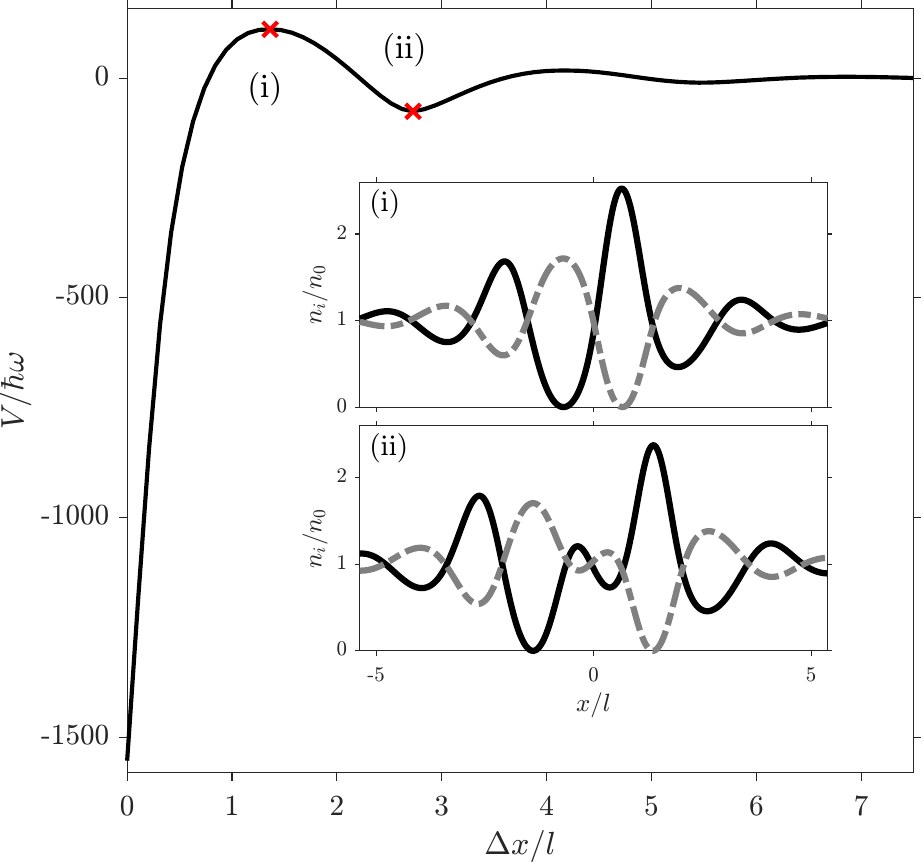}
	\caption{Inter-soliton potential versus soliton separation for a binary dipolar condensate.
	Three minima are present, corresponding to three regions where bound states can form.
	The insets show the density of component 1 (solid) and component 2 (dashed) of soliton pairs with separations $\Delta x/l\approx1.35$ and $\Delta x/l\approx2.7$, corresponding to a maximum (i) and a minimum (ii) in $V$, respectively.
	Contact interaction parameters: $\{a_{11},a_{12},a_{22}\}/a_0 = \{140,105,140\}$. }
	\label{Fig:BSOrigin}
\end{figure}

To elucidate the origin of multiple minima in the inter-soliton potential $V(\Delta x)$, we examine the case $a_{12}/a_0=105$ in Fig.~\ref{Fig:BSOrigin}, where minima occur at $|\Delta x| /  l \approx[0,2.7,5.4]$.
In the presence of spin rotons, a single soliton induces a spatially oscillating spin-density profile (recall Figs.~\ref{Fig1:Dens} and \ref{Fig2:Dispersion}) that acts as a polarized background.
If a second soliton is placed at a location where this background aligns with its polarization distribution, only minor adjustments of the spin texture are required [see Fig.~\ref{Fig:BSOrigin}(i)], lowering the energy cost compared to a homogeneous background (yielding $E<0$, i.e., $V>0$).
Conversely, if the background is anti-aligned, large spin-density rearrangements are needed, raising the energy cost ($E>0$, i.e., $V<0$).
In this case, the configuration shown in Fig.~\ref{Fig:BSOrigin}(ii) stabilizes a bound stationary state at $\Delta x /  l \approx 2.7$, matching the roton wavelength $\lambda_\text{rot} /l\approx2.7$.
Although this may seem counterintuitive, the restoring force is directed towards maxima of $E$ (minima of $V$) due to the soliton's negative effective mass [see Eq.~(\ref{Eq:ISP})].

When approaching the roton instability, $a_{12}^\text{c}/a_0\approx110$, the decay length of the spatial spin density oscillations around the soliton core grows, so too does the number of alternating peaks and troughs of the inter-soliton potential energy.
For soliton pairs with like sign, minima in $V(\Delta x)$ develop at $|\Delta x| \approx (I+1/2)\lambda_\text{rot}$, whereas for opposite-sign pairs they occur at $|\Delta x| \approx I\lambda_\text{rot}$, for $I=0,1,2,\dots$\,. 
Correspondingly, maxima in $V(\Delta x)$ occur at $|\Delta x| \approx I\lambda_\text{rot}$ for like-sign solitons, and at $|\Delta x| \approx (I+1/2)\lambda_\text{rot}$ for opposite-sign solitons.

Since spin rotons are predicted across a broad range of dipolar mixtures \cite{kirkby2024excitations}, our arguments imply that multiple bound states are expected not only in dipolar-nondipolar mixtures but also more generally.
Additional simulations in Appendix~\ref{Appendix:OtherMixtures} corroborate this, showing such bound states for $\mu_1/\mu_2 = -1$ and $\mu_1/\mu_2 \approx 0.2$.

\subsection{Soliton collisions}

True solitons can, in principle, emerge unscathed from collisions.
However, the presence of dipolar interactions breaks integrability, so perfectly elastic collisions are not expected.
Nevertheless, dark-antidark solitons may still survive collisions and thus serve as a probe of the inter-soliton potential.

\begin{figure}[h!]
	\centering
	\includegraphics[width=0.48\textwidth]{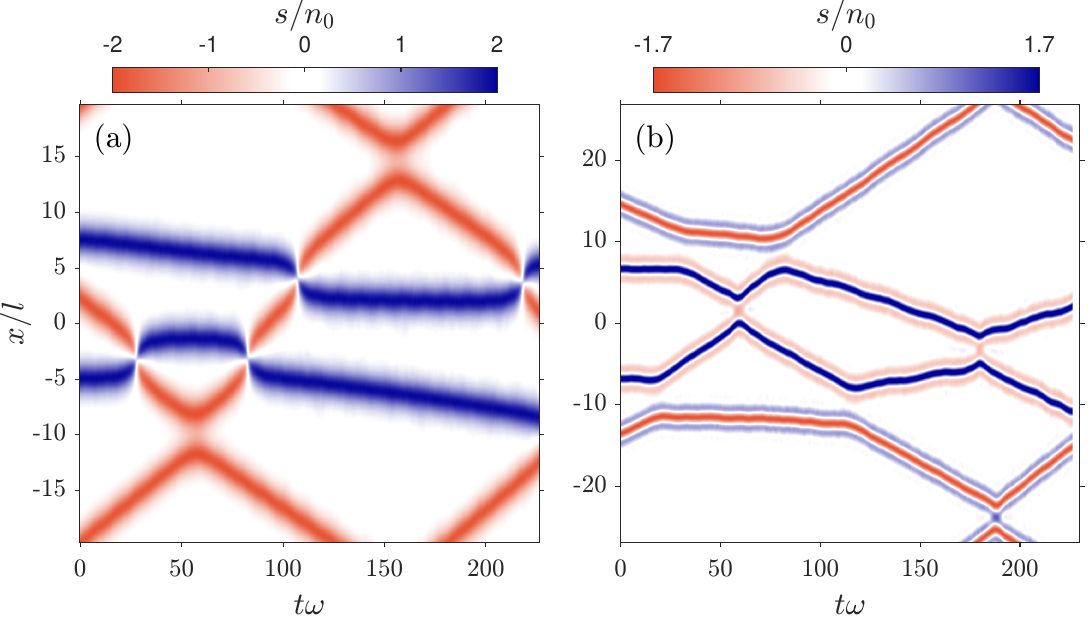}
	\caption{Unbound collisional dynamics involving two initially nonmoving positive-core (blue) and two initially moving negative-core (red) dark-antidark solitons, shown (a) without and (b) with dipolar interactions. Spin density is shown as a function of position and time. Contact interaction parameters: (a) $\{a_{11},a_{12},a_{22}\}/a_0 = \{100,95,100\}$; (b) $\{a_{11},a_{12},a_{22}\}/a_0 = \{140,95,140\}$. }
	\label{Fig:Coll}
\end{figure}

Figure \ref{Fig:Coll} shows four dark-antidark solitons in each panel---two slowly moving negative-core solitons and two stationary positive-core solitons---in (a) the nondipolar reference system and (b) a binary dipolar condensate.
In the nondipolar case, like-sign solitons bounce off each other, whereas opposite-sign solitons pass through (as predicted in Ref.~\cite{rohrs2025magnetic}).
With dipolar interactions present, however, they always bounce, independent of relative sign.
The bouncing arises from the periodic maxima in the inter-soliton potential $V$, which act as effective barriers.
Crucially, universal low-velocity bouncing occurs because these maxima exist regardless of the soliton signs (see Sec.~\ref{Sec:BSOrigin}).
For like-sign solitons, these maxima occur at integer multiples of $\lambda_\text{rot}$; for opposite-sign solitons, half-integer multiples (see Fig.~\ref{Fig:BSOrigin}).
Consequently, for the collisions shown in Fig.~\ref{Fig:Coll}(b), the minimum separation before bouncing differs: $\Delta x\gtrsim\lambda_\text{rot}$ for like-sign collisions and $\Delta x\gtrsim1.5\lambda_\text{rot}$ for opposite-sign collisions, suggesting that in both cases the second maxima are relevant at these velocities.
These distinct collisional properties---especially when contrasted with the nondipolar case---could provide direct experimental evidence of the spin roton.

\section{Conclusion}

We investigated the formation of multiple dark-antidark soliton bound states in two-component dipolar BECs. Using Bogoliubov theory, we derived the dispersion relation and linked the emergence of spin rotons to strong spatial spin-density oscillations around the soliton cores and modulations of the effective inter-soliton potential, leading to the multiple bound states at distinct separations.
Soliton collisions revealed that maxima in the inter-soliton potential act as effective barriers, causing solitons to bounce regardless of relative sign.
This behavior stands in sharp contrast to the nondipolar case---where opposite-sign solitons always pass through one another---thus providing a realistic pathway to probe and validate spin rotons in experiments.

Our results remain robust across a broad range of parameters, including densities well below those presented here. Moving toward higher densities, approaching the 3D-cigar regime, could test the persistence and stability of soliton bound states, and may uncover new links between roton physics and nonlinear excitations.\\

{\it Acknowledgements:---}
The authors acknowledge T. Billam, T. Bland, F. Ferlaino, L. B. Giacomelli, A. Madhusudan, N. Masalaeva, N. Parker, E. Poli, C. Qu and P. Senarath Yapa for insightful discussions.
This research was funded by the Austrian Science Fund (FWF) [10.55776/P36850].


%

\appendix

\section{Additional dipolar mixtures}
\label{Appendix:OtherMixtures}

\begin{figure}
	\centering
	\includegraphics[width=0.48\textwidth]{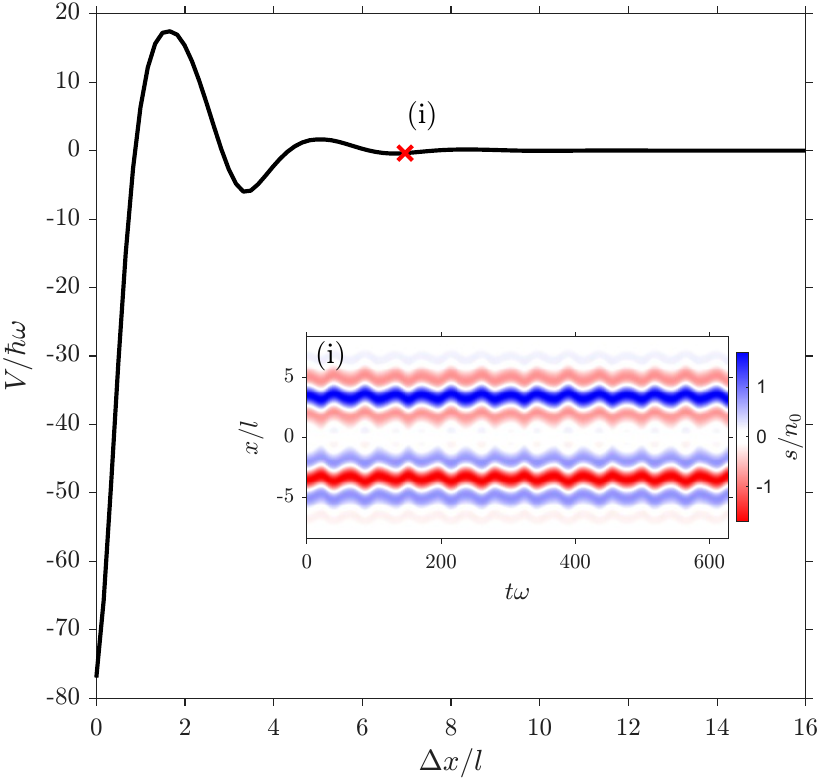}
	\caption{Inter-soliton potential versus soliton separation for a binary dipolar condensate with $\mu_1=-\mu_2$.
	Four minima are present ($x_1=0,x_2\approx 3.3,x_3\approx6.7, x_4\approx10.1$), corresponding to four regions where bound states can form.
	The inset shows example dynamics of the third bound state in an excited configuration, with the initial state position indicated as a cross in the main figure.
	Contact interaction parameters: $\{a_{11},a_{12},a_{22}\}/a_0 = \{140,35,140\}$.
		}
	\label{Fig:6}
\end{figure}

To demonstrate the generality of our findings to the case where both $\mu_1, \mu_2 \neq 0$, we consider two additional dipolar configurations. 
In the first, we choose $\{\mu_1, \mu_2\} = \{9.93, -9.93\} \mu_{\rm B}$ and an average per-component density $n_0 = 133~\text{atoms}~\mu\text{m}^{-1}$. Unless otherwise specified in the captions, the remaining parameters are the same as in Sec.~\ref{Numerics}.
Figure \ref{Fig:6} shows the inter-soliton potential between opposite-sign solitons, with four minima present, each representing a different bound state. Figure \ref{Fig:6}(i) shows an example of one such bound state in an excited configuration, oscillating about the third minimum in Fig.~\ref{Fig:6}. 
Importantly, both the inter-soliton potential and bound state behavior are qualitatively similar to the results presented in Fig.~\ref{Fig:Pot}, despite the vastly different combination of dipole-dipole interaction parameters.

\begin{figure}
	\centering
	\includegraphics[width=0.48\textwidth]{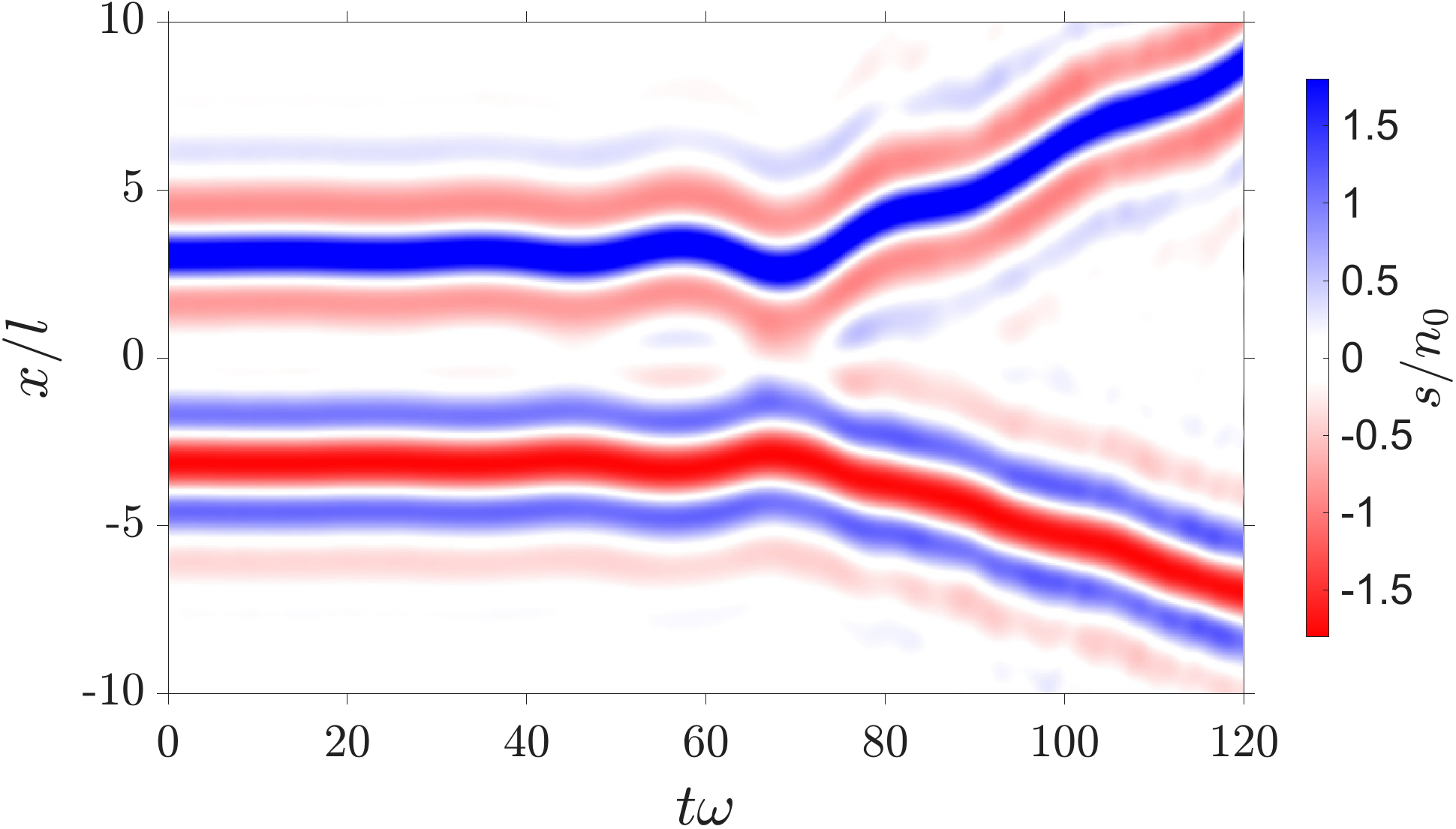}
	\caption{Example of a bound state for $\{\mu_1, \mu_2\} = \{9.93, 2\} \mu_{\rm B}$.
	This particular state becomes unbound at $t\omega \approx 70$. 
	Contact interaction parameters: $\{a_{11},a_{12},a_{22}\}/a_0 = \{140,121,140\}$.
	}
	\label{Fig:muB10muB2}
\end{figure}

In the second configuration, we set $\{\mu_1, \mu_2\} = \{9.93, 2\} \mu_{\rm B}$. Figure \ref{Fig:muB10muB2} shows an example of a bound state that remains stable until $t\omega \approx 70$.
While these results may at first seem to suggest that a smaller difference between dipole moments may lead to shorter bound-state lifetimes, we also sometimes observed relatively short-lived bounds states for dipolar-nondipolar mixtures, hence further investigations are required to properly characterize lifetimes.

\section{Bogoliubov theory}
\label{Appendix:BdG}
This derivation of the dispersion relations for elementary excitations is similar to the one in Ref.~\cite{kirkby2023spin} for antidipolar mixtures.

We look for solutions of Eq.~(\ref{Eq:GPE}) of the form:
\begin{equation}
    \psi_i(x,t)=\left(\sqrt{n_i}+\lambda \left[u_i(x)e^{-i\epsilon t/\hbar}-v_i^*(x)e^{i\epsilon^* t/\hbar}\right]\right)e^{-i\mu_it/\hbar},
    \label{eq.state}
\end{equation}
where the condensate is perturbed via a small parameter $\lambda$ around the average density $n_i$ and $u_i$ and $v_i$ are the Bogoliubov amplitudes describing the perturbation.
Inserting Eq. (\ref{eq.state}) into Eq.~(\ref{Eq:GPE}) and keeping only terms up to linear order in $\lambda$, we can find the coupled equations by collecting terms evolving in time with $e^{-i\epsilon t/\hbar}$ and $e^{i\epsilon^* t/\hbar}$. Moving to momentum space $\big[\mathcal{F}_x[a(x)]=\Tilde{a}(k)\big]$ we get
\begin{align}
        \epsilon \Tilde{u}_i(k)=\frac{\hbar^2k^2}{2m} & \Tilde{u}_i(k)+\frac{\sqrt{n_i}}{2\pi l^2}\sum_jg_{ij}\sqrt{n_j}\left[\Tilde{u}_j(k)-\Tilde{v}_j(k)\right] \notag \\
       + ~ \sqrt{n_i} & \sum_jU_{ij}(k)\sqrt{n_j} \left[ \Tilde{u}_j(k)- \Tilde{v}_j(k)\right]
\end{align}
and
\begin{align}
           \epsilon \Tilde{v}_i(k)=-\frac{\hbar^2k^2}{2m} & \Tilde{v}_i(k)+\frac{\sqrt{n_i}}{2\pi l^2}\sum_jg_{ij}\sqrt{n_j}\left[\Tilde{u}_j(k)-\Tilde{v}_j(k)\right] \notag \\
          + ~ \sqrt{n_i} & \sum_jU_{ij}(k)\sqrt{n_j} \left[ \Tilde{u}_j(k)- \Tilde{v}_j(k)\right].
\end{align}

The coupled equations can be written as a matrix equation of the form $\epsilon \textbf{w}=M\textbf{w}$, where $\textbf{w}=\{\Tilde{u}_1,\Tilde{u}_2,\Tilde{v}_1,\Tilde{v}_2\}$.
The eigenvalues can be analytically calculated and the two positive-norm energy branches are then given by
\begin{equation}
    \epsilon_{\pm}^2(k)=\frac{\epsilon_1^2+\epsilon_2^2}{2}\pm\frac{1}{2}\sqrt{(\epsilon_1^2-\epsilon_2^2)^2+4G_{12}G_{12}\frac{\hbar^4k^4}{m^2}},
    \label{eq.modeenergies}
\end{equation}
where the quasiparticle mode energies are
\begin{equation}
    \epsilon_i^2(k)=\frac{\hbar^2k^2}{2m}\left(\frac{\hbar^2k^2}{2m}+2G_{ii}(k)\right),
\end{equation}
with
\begin{align}
    G_{ij}(k)=\sqrt{n_in_j} \left( \frac{g_{ij}}{2\pi l^2}+U_{ij}(k) \right) . \label{Eq:GijApp}
\end{align}
Equations (\ref{eq.modeenergies}-\ref{Eq:GijApp}) appear in the main text as Eqs.~(\ref{Eq:BdG}-\ref{Eq:Gij}).

\vspace{1cm}

\section{Soliton effective mass}
\label{Appendix:EMS}
\begin{figure}[h!]
	\centering
	\includegraphics[width=0.48\textwidth]{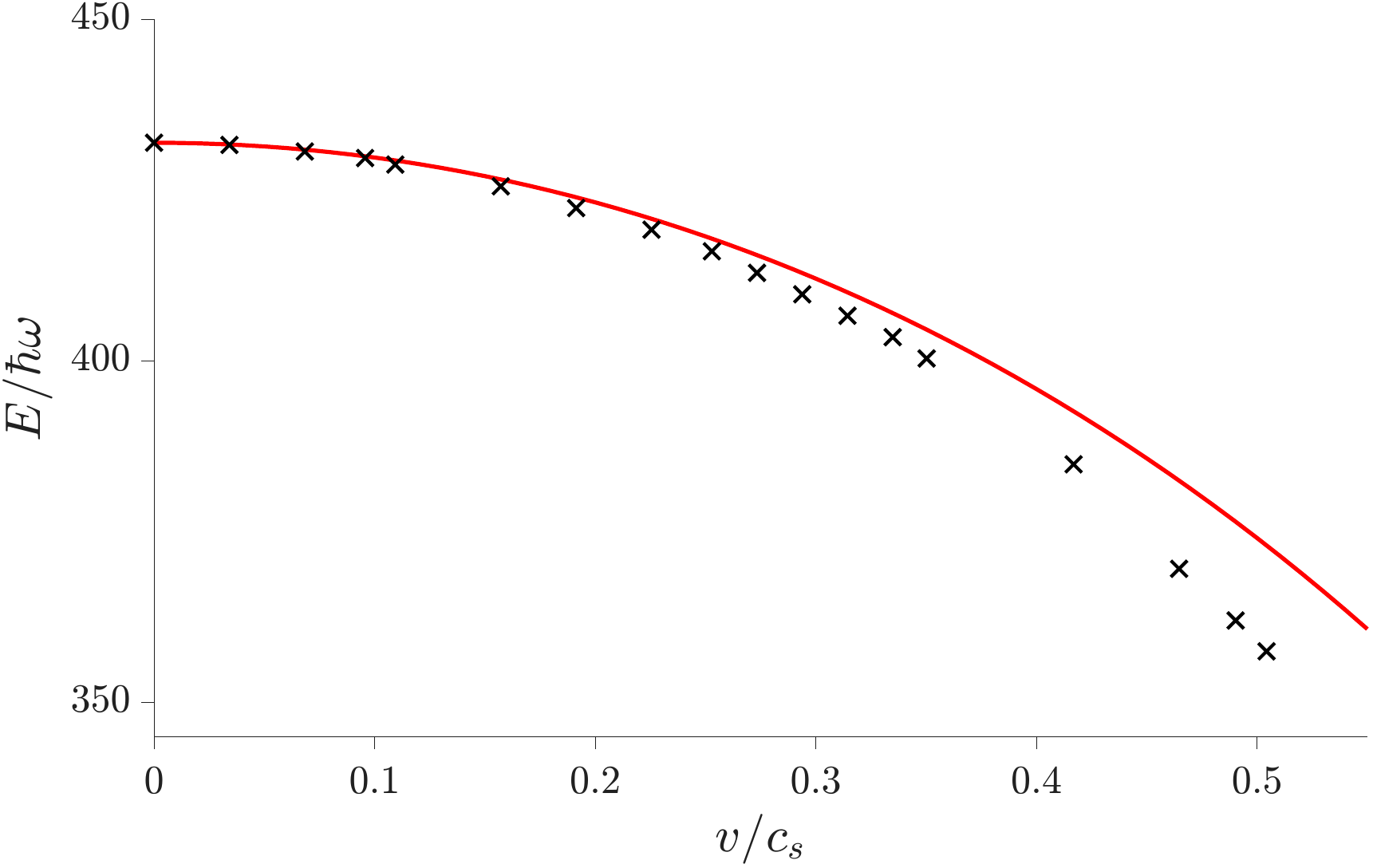}
	\caption{Energy vs velocity for a dark-antidark soliton in a dipolar-nondipolar mixture. Black crosses show the numerical predictions, while the red solid line shows the analytic prediction for a magnetic soliton in a binary nondipolar BEC. 
	Contact interaction parameters: $\{a_{11},a_{12},a_{22}\}/a_0 = \{140,95,140\}$.
	}
	\label{Fig:8}
\end{figure}

Here, we investigate the sign of the effective mass of dark-antidark soltions in binary dipolar BECs.
In Fig.~\ref{Fig:8}, we numerically calculate the energy $E$ (black crosses) of a single dark-antidark soltion as a function of its velocity $v$.
For reference, we also plot the analyticaly predicted energy of a magnetic soliton in a binary nondipolar BEC (red line) \cite{qu2016magnetic}
\begin{equation}
E_{\rm{ms}}(v)=\alpha\sqrt{1-v^2/c_{\rm{s}}^2},
\end{equation}
where $c_{\rm{s}}$ is the spin speed of sound and we choose $\alpha$ such that $E_{\rm{ms}}(v=0)=E(v=0)$.
We can see that $E(v)$ decreases as $v$ increases, which indicates a negative effective mass according to the following expression, \cite{qu2017magnetic}
\begin{equation}
m_{\rm{eff}}=\frac{1}{v_{\rm{s}}}\frac{dE}{dv_{\rm{s}}}.
\end{equation}

\end{document}